\documentclass[preprint2]{aastex}

\begin{document}

\title{The Evolution of Protoplanetary Disk Edges}

\author{Peggy Varniere, A.~C. Quillen \& Adam Frank}
\affil{Department of Physics and Astronomy,
University of Rochester, Rochester, NY 14627;\\
{\it pvarni, aquillen, afrank@pas.rochester.edu
}
}

\begin{abstract}
We investigate gap formation in gaseous protostellar disks by a
planet in a circular orbit in the limit of low disk viscosity.
This regime may be appropriate to an aging disk after the epoch of
planet formation. We find that the distance of planet to the gap outer boundary
can be between the
location of the $m=2$  and $m=1$ outer Lindblad resonances.
This distance is weakly dependent upon both the planet's mass and disk
viscosity. We find that the evolution of the disk edge takes place
on two timescales.  The first timescale is set by the spiral
density waves driven by the planet. The second timescale depends
on the viscosity of the disk. The disk approaches a state where
the outward angular momentum flux caused by the disk viscosity
is balanced by the dissipation of spiral density waves which are
driven at the
Lindblad resonances.  This occurs inefficiently however because of
the extremely low gas density near the planet. We find that the
distance between the planet and the peak density at the
disk outer edge is only weakly dependent on the viscosity and planet
mass, however the ratio of the gas density near the planet to that
in the disk (or the slope of density along the disk edge) is
strongly dependent upon both quantities.  We find that the disk
density profile along the edge scales approximately with disk viscosity
divided by the square of the planet mass.  We account for this
behavior with a simple scenario in which the dissipation of
angular momentum from the spiral density waves is balanced
against diffusion in the steep edge of the disk.

\end{abstract}

\section{Introduction}

The discovery of extra-solar planetary systems \citep{mayor} has driven a
flood of renewed interest in planetary formation and disk evolution models.
The focus of many models has been on mechanisms that couple the planets
to the evolution of protostellar disks,
the tidal interactions between a planet
and a protoplanetary disk \citep{artymowicz94,ogilvie}
and resulting planetary migration \citep{lin86,trilling,nelson,murray}.
Gaps and holes in protostellar disks
in some cases have been detected
through the study of spectral energy distributions \citep{rice,calvet,koerner}
and direct imaging \citep{jay,augereau,weinberger,grady}.
Since structure in protostellar disks can arise through
the gravitational interaction between the disk and planets, detailed
studies of these disks may allow us to place constraints on the properties
of the planets in them and on the evolution of these systems.

The formation of gaps in gaseous or planetesimal disks
has a long history of theoretical study.
It has long been postulated that an external time dependent
gravitational potential, such as from a satellite or planet,
could resonantly drive waves at Lindblad resonances (LRs) in a gaseous
or planetesimal disk \citep{GT78, GT79, lin79, pp3, ward}.
Because these waves carry angular momentum, they govern the
way that planets open gaps in disks  (e.g., \citealt{bryden,artymowicz94})
and can result in the orbital migration of planets
(e.g., \citealt{nelson}).
Previous works have derived and numerically verified
conditions for a planet to open a gap in a gas disk
(e.g., \citealt{bryden}).
These studies focused on when gaps occur and
explored the accretion rate through the gap as material was
cleared to either side.  Emphasis was also placed on
the relationship between gap properties, accretion and orbital
migration rates (e.g., \citealt{kley,nelson}).

However, little numerical work has focused on the expected sizes
and depths of the gaps in different physical regimes.
Such studies
are warranted because, as discussed above, gaps and holes in
protoplanetary disks are detectable
and so will in principle allow us to
place constraints on the properties of the planets that  are
responsible for maintaining them.

The physical properties of protoplanetary disks should vary as the disk ages.
We parametrize the disk viscosity, $\nu$, with a Reynolds number,
${\cal R} \equiv r^2 \Omega/\nu$ where $\Omega$ is the angular
rotation rate of a particle in a circular orbit around the star
and $r$ is the radius.
Using an $\alpha$ form for the viscosity,
\begin{equation}
{\cal R }
= \alpha^{-1} \left( { v_c \over c_s }\right)^2
= \alpha^{-1} \left( { r \over H }\right)^2
\end{equation}
where $c_s$ is the sound speed, $H$ is the vertical scale height
of the disk and $v_c = r \Omega$ is the velocity of a particle in
a circular orbit. We expect that the sound speed and vertical
scale height of a cooling gaseous disk will drop with time,
increasing the Reynolds number. For a planetesimal disk, the
effective viscosity depends upon the opacity, $\tau$, (or area
filling factor) of the disk with $\alpha \sim {\tau \over 2}$ in
the limit of low $\tau$ (e.g., \citealt{M+D} and citations
within). As planetesimals coagulate or are scattered out of the
disk, we expect a decrease in the viscosity and so an increase in
the Reynolds number.

When viscous forces are overcome by the angular momentum flux
driven by density waves, a gap in the disk near the planet opens.
So as protoplanetary disks age (and their viscosity drops) we
expect that gaps caused by planets will widen and accretion onto
these planets will cease \citep{bryden,kley}. Because of the
longer timescale associated with the end of accretion, it may be
that forthcoming surveys will be dominated by partially cleared
disks. The advent of new instruments such as SIRTF, ALMA, SMA and VLTI
make a study of this limit particularly interesting and relevant
(e.g., \citealt{wolf}).

In this paper we numerically investigate the properties of gaps,
exploring a larger range of parameter space than have previous
studies.  In particular we extend our simulations toward the low
viscosity limit, which may govern a longer timescale in the
evolution of protoplanetary disks.  We critically review our
results in the context of previous theories of gap formation and
attempt to extend these models to embrace our new results.
In a follow up paper we will explore the observational properties of
our simulated disks \citep{varni04}.

The structure of our paper is as follows.  In section 2 we review
the theory of gap formation. In section 3 we present our
numerical methods and the input conditions for our models.
In section 4 we present our results and in section 5 we present a
critical discussion of the results along with attempts to extent
previous gap formation models. A discussion and conclusion
follows in section 6.

\section{Previous Theory}
Linear theory has been developed to predict the angular momentum
flux driven by spiral densities waves exited by a planet,
\citep{GT78,donner,lin79,artymowicz}. This theory has
successfully been used to predict the formation of gaps seen in
numerical simulations \citep{pp3,bryden}. If a gap is opened then
the angular momentum flux resulting from spiral density waves can be
balanced by the angular momentum flux caused by the viscosity of
the disk. To theoretically account for these phenomena,
we require expressions
for the angular momentum flux due to both processes.

As have previous works, we assume that the perturbing
gravitational potential from the planet can be expanded in Fourier
components
\begin{equation}
\Phi(r,\theta, t) = \sum_m W_m(r) \exp\left( im(\theta - \Omega_p t)\right)
\end{equation}
where $\Omega_p$ is the angular rotation rate of the pattern, and
$r, \theta$ are polar coordinates in the plane containing the
planet. Here $\Omega_p$ is the angular rotation rate of a planet
which we assume is in a circular orbit. The torque $T_m$ exerted at
the $m$-th Lindblad resonance into a disk of surface density
$\Sigma$ was found to depend solely on the radial profile of the
perturbing potential and be independent of the process of
dissipation providing that the waves are completely dissipated
\citep{GT78,donner,lin79,artymowicz},
\begin{equation}
\label{T_m}
T_m = - m \pi^2 \Sigma { |\Psi_{GT}|^2 \over r d D_m /dr}
\end{equation}
where
\begin{equation}
\Psi_{GT} = r {dW_m \over dr} + {2 m \Omega \over m(\Omega - \Omega_p)} W_m
\end{equation}
and
\begin{equation}
D_m = \kappa^2 - m^2(\Omega - \Omega_p)^2.
\end{equation}
Here $\Omega$ and $\kappa$ are the angular rotation
rate and epicyclic frequency of an unperturbed particle in a
circular orbit and are functions of radius.
The above torque results from $m$-armed waves driven at the
$m$th Lindblad resonance which is located at a radius determined by
$D_m(r_{LR})=0$.
Equation(\ref{T_m}) is evaluated at this radius.

For a planet in a circular orbit, the Fourier components of the potential
exterior to the planet are
\begin{equation}
W_m = - { G M_p \over  r} b^m_{1/2}(r_p/r )
\end{equation}
where $b^m_{1/2}$ is the Laplace coefficient (e.g., see \citealt{M+D})
and $r_p$ is the radius of the planet from the central star.
We write equation (\ref{T_m}) in the form
\begin{equation}
T_m = -f_m \Sigma r_p^4 \Omega_p^2 \left({q \over 10^{-3}}\right)^2
\label{Tmfm}
\end{equation}
where $f_m$ are unitless constants,
$q \equiv M_p/M_*$ is the mass ratio of the planet to the central star,
and $\Sigma$ is evaluated at the Lindblad resonance.
Evaluating equation(\ref{T_m}) using summations for
the Laplace coefficients (e.g., \citealt{M+D}) we find that
$f_1= 10^{-5.1}$,
$f_2 = 10^{-4.3}$, $f_3=10^{ -3.7}$ and $f_4 = 10^{-3.4}$. These values
will be used in section 4 in our discussion on clearing timescales
and in section 5 in our discussion on the edge profile.

If the planet is sufficiently massive and the disk viscosity
sufficiently low, the torque caused by the excitation of
spiral density waves will overcome the inflow due to viscous
accretion in the disk and a gap will open. The angular momentum
flux or torque transferred through a radius $r$ in a Keplerian
viscous disk with constant viscosity and density is given by
\begin{equation}
\label{T_nu}
T_\nu = 3 \pi \Sigma \nu r^2 \Omega.
\end{equation}
Previous estimates of gap opening criteria have applied
equation(\ref{T_m}) in the limit for large $m$ so that the
resonances closest to the planet are taken into account
\citep{lin79,bryden}. In these cases the total torque resulting
from summing the resonant terms from each outer Lindblad resonance
(OLR) within a distance
$\Delta$ from the planet yields
\begin{equation}
T \approx 0.23 q^2 \Sigma r_p^2 \Omega^2 \left( \frac{r_p}{\Delta} \right)^3
\label{tm}
\end{equation}
where $r_p$ is the radius of the planet \citep{pp3}. In what
follows we take $\Delta$ to be the distance from the planet to
the gap outer boundary.
The expression above is valid when the gap
is larger than the maximum value of the disk scale height ($\Delta
> H$). When the torque from the dissipation of the outward going density waves is
balanced by the torque from viscous accretion (equation
\ref{T_nu}) one can derive a criterion for gap formation, {\rm
i.e.} $q > 40/{\cal R}$ \citep{bryden}.

Once a gap is opened, the torque from viscous accretion must be
balanced by the total torque from the spiral density waves driven
at Lindblad Resonances exterior to the gap. As the mass of
the planet increases and the viscosity decreases, the gap edges
moves away from the planet so that the total torque contains fewer resonant
terms.  Past the $m=2$ OLR at $r \sim 1.3 r_p$ only the weaker $m=1$ OLR at
$r\sim 1.6 r_p$ can drive spiral density waves.   Previous works have
suggested that the location of the $m=2$ OLR resonance would set the radius
of disk edges in low viscosity disks harboring large mass planets
\citep{pp3}.  It is possible that as the gap is cleared, gas could
accumulate near this resonance \citep{pp3}.  Alternatively, spiral
density waves could be driven at a LR via a `virtual
effect' as long as the resonance was within a particular distance
of the disk edge.  This argument has been used to provide a
possible explanation for $m=2$ structure seen in simulations of
binaries (e.g., \citealt{pp3,lin79,savonije}).
A large mass ratio,
such as would be found in a stellar binary system,
or extremely small disk viscosity,
is required to open a gap out to
the $m=1$ OLR in a circumbinary disk \citep{artymowicz94}.

\begin{figure*}
%
\plotone{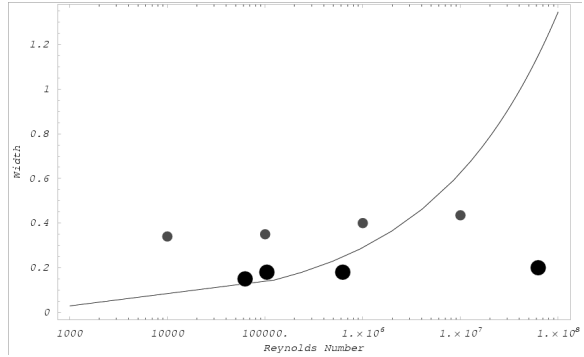}
\figcaption{Disk width vs Reynolds number. The $y$-axis
shows the distance from the planet to the outer disk edge.
The $x$-axis shows the Reynolds number.  The solid line shows the prediction
based on equation \ref{delscale}. Large points are measured from
simulations displayed by \cite{bryden}.
Small points are taken from simulations shown in this paper with planet
mass ratio $q=10^{-3}$. Note that the definition of the gap width differs
slightly for the two points (see section 4.1). This plot
demonstrates that simple models (such as based on equation \ref{delscale})
fail to predict the dependence of gap width on viscosity.
\label{adam}}
\end{figure*}

The balance between torque from spiral density waves and $T_\nu$
should allow one to derive a scaling
relation for the gap width.   If we naively use equation (\ref{tm})
and assume a Keplerian rotation profile, we find
\begin{equation}
\Delta = 0.29 q^{2/3}{\cal R}^{1/3} r_p. \label{delscale}
\end{equation}
While previous numerical studies have confirmed the condition for
opening a gap, a general scaling relation to predict the gap width has not
yet been developed. However, it is possible to extract some
information from published data in order to probe the
scaling of gap width with planet mass and disk viscosity.  In
Figure \ref{adam} we present a plot of equation(\ref{delscale})
along with data from \citet{bryden} as well as data from our own
simulations. The points by \citet{bryden} were taken for their
$q=10^{-3}$ simulations and the gap widths were measured from
their density profile plots.  The gap width was taken to be the
distance from the planets orbital radius to the location along the
outer gap wall where the density is half of that outside the gap.
Reynolds numbers were computed using their quoted values of the
viscosity parameter $\alpha_{ss}$ and
\begin{equation}
\nu = \alpha_{ss} ( H/r )^2 r^2 \Omega.
\end{equation}
For the zero viscosity case of \citet{bryden}, we use their quoted
value of the numerical viscosity yielding $\alpha_{ss} = 10^{-5}$.

Inspection of Figure \ref{adam} shows that the scaling relation
given in equation (\ref{delscale}) produces the correct order of
magnitude gap width.   Equation(\ref{delscale}), since it is based
upon a high $m$ integration, is correct for lower planet masses
and higher disk viscosities, providing an explanation for the
correct positions of the points in Figure \ref{adam} on the high
viscosity end. Given that the gap edge is not sharp, there is some
ambiguity in the definition of $\Delta$ and hence in the exact
comparison between equation(\ref{delscale}) and the simulations.
We take up this point again in section 4. In spite of this
uncertainty, the failure of the scaling relation to predict the
gap width across a wide range of Reynolds numbers is clearly seen
in the trend shown by the simulations of \citet{bryden}. The gap
width increases more slowly than predicted by
equation(\ref{delscale}) and is off by a factor of $2$ or more at
low viscosity. Again this is consistent with the fact that
equation(\ref{delscale}) does not appropriately estimate gap
widths when the edge of the disk is near the lower order Lindblad
resonances.
It is difficult to draw firm conclusions from this data since
\citet{bryden} do not explore the relation across a uniform (in
log) sampling of viscosities. Figure \ref{adam} also shows data
from our simulations with a more complete sampling of viscosity
and these confirm the weak dependence of gap width on
Reynolds number.

Thus it appears that while previous approaches can predict when a
gap forms they do not capture the full physics of the gaps. In
particular some aspect of the scenario in which the torque from
waves driven at the OLR is balanced by viscous accretion (as
discussed by \citealt{bryden}) must be incomplete. This is not a
trivial point as observational diagnostics from platforms such as
SIRTF and ALMA may determine gap widths to less than 10\% 
\citep{wolf}. Since many of the observable systems may be
appropriate for the low viscosity limit, understanding the proper
scaling there will be critical to using observations of gap widths
and disk edges to infer protoplanet properties.

\section{Numerical Code Description}

To investigate the evolution of gaps opened by a protoplanet, we
ran a set of $2$D hydrodynamical simulations using the code
developed by \citet{masset02,masset03}. This code is an Eulerian
polar grid code with a staggered mesh and an artificial second
order viscous pressure to stabilize the shocks (see also
\citealt{stone}). The hydrocode allows tidal interaction between
one or more planets and a 2D non-self-gravitating gaseous disk,
and is endowed with a fast advection algorithm that  removes the
average azimuthal velocity for the Courant timestep limit
\citep{masset00}.  The simulations are performed in the
non-inertial non-rotating frame centered on the primary star
(similar to a heliocentric frame). The
outer boundary does not allow either inflow or outflow, so it must
be located sufficiently far from the planet to ensure that spiral
density waves are damped before they reach it. This is facilitated
by adopting a logarithmic grid in radius. The grid inner boundary
only allows material to escape so that the disk material may be
accreted on to the primary star.


The code is scaled so that the unit length is $1$ AU which is the
initial position of the protoplanet, the unit mass is $1\ M_\odot$,
which is the mass of the central star,
and time is given in units of one orbital period (one year)
for an object in a circular orbit at the initial location of the planet.

The runs we are presenting here are made with a resolution of
$N_r = 150$ and $N_\theta = 450$.
The grid physical size spans radii between $0.25$ AU and $6.0$AU.
The  viscosity is parametrized with the Reynolds number
${\cal R} \equiv r^2 \Omega/\nu$, where $\cal R$ is assumed to be constant
with radius.  The disk aspect ratio $H/r = 0.04$ is uniform and constant where
$H$ is the vertical scale height of the disk.
The sound speed of the gas is set from the disk aspect ratio.
The planet was initially embedded in a disk
with density profile
$\Sigma(r) = 10^{-5} \left({r \over R_{min}}\right)^{-1.5}$ where $R_{min}$
is the radius of the inner edge of the grid.
Table \ref{tab:run} shows the characteristics of each run we will discuss here.

\section{Results}

\begin{figure*}
\plotone{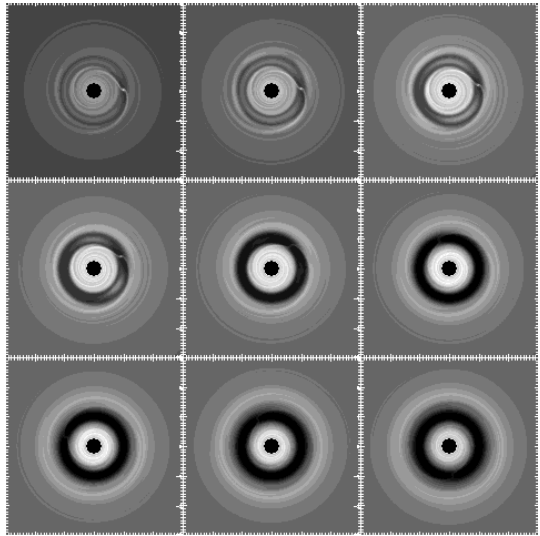}
\figcaption{
2-D simulation of a $q=10^{-3}$
(approximately Jupiter mass)
planet opening a gap in a gas disk with Reynolds number ${\cal R} =2 \times 10^6$.
The gas density is shown at
times 10, 20, 50, 100, 200, 500, 1000, 2000, and 3000 after the beginning
of the simulation.   Times are given
in units of rotational period of the planet around the star.
The top three panels show the initial clearing of the gap on a fairly
short timescale.  The bottom three panels show the longer timescale
behavior.
On long timescales the gap opened by the planet is
between the $m=2$ and $m=1$ OLR.
\label{protodisk.ps}}
\end{figure*}

Figure \ref{protodisk.ps} show snapshots of the density for a
typical simulation, namely $q = 10^{-3}$ (corresponding to a Jupiter mass planet) and a Reynolds number
${\cal R} = 2 \times 10^6$. From top to bottom and left to right we show the density
distribution at times
$t= 20, 50, 100, 200, 500, 1000, 2000 $ and 3000
(in units of the planet's orbital period).
We see that a gap is cleared
relatively quickly during the first hundred rotation periods.
During this time period a prominent 2-armed structure is seen in the density
 distribution.
At later times, the density near the planet drops and the
amplitude of the spiral density waves decreases. An asymptotic
state is approached on longer timescales when the edge of the disk
is between the $m=2$  and $m=1$ OLRs.
We see that the gas interior to the planet slowly accretes
onto the star while that exterior to the planet slowly accretes
inward piling up near the edge of the disk, outside the planet.

The shape of the disk edge at latter times is better seen on Figure
\ref{fig:gap_def}a which shows the density profile of the
inner region of the disk for the same sampled times. Figure
\ref{fig:gap_def}a also shows 3 radial functions that can be
used as metrics to measure $\Delta$, the ``outer'' width of the gap or
the distance between the planet and the gap outer edge.


\subsection{Measuring the distance between the planet and the edge of the disk}

\begin{figure*}
\epsscale{0.60}
\plotone{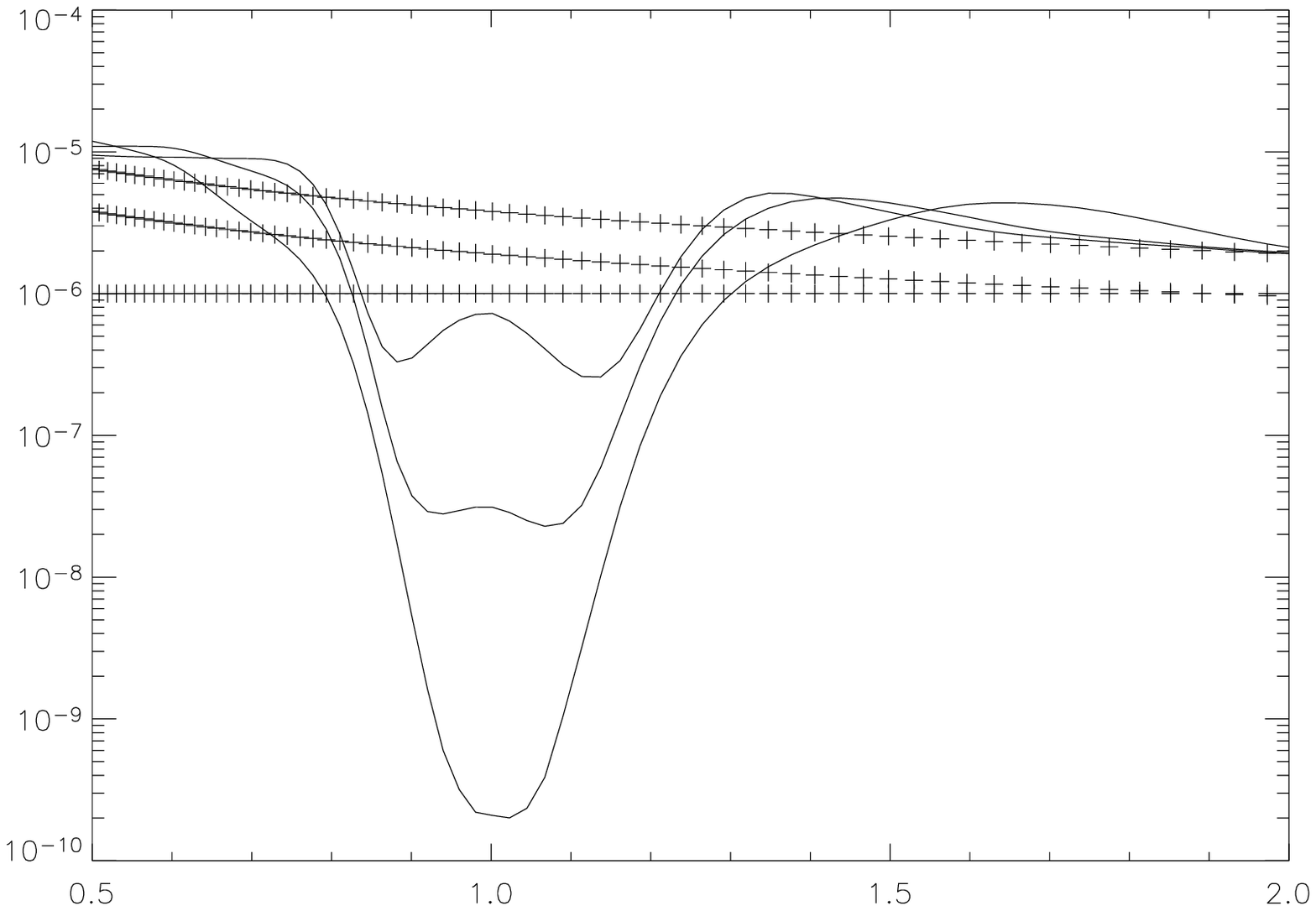}
\epsscale{0.50}
\plotone{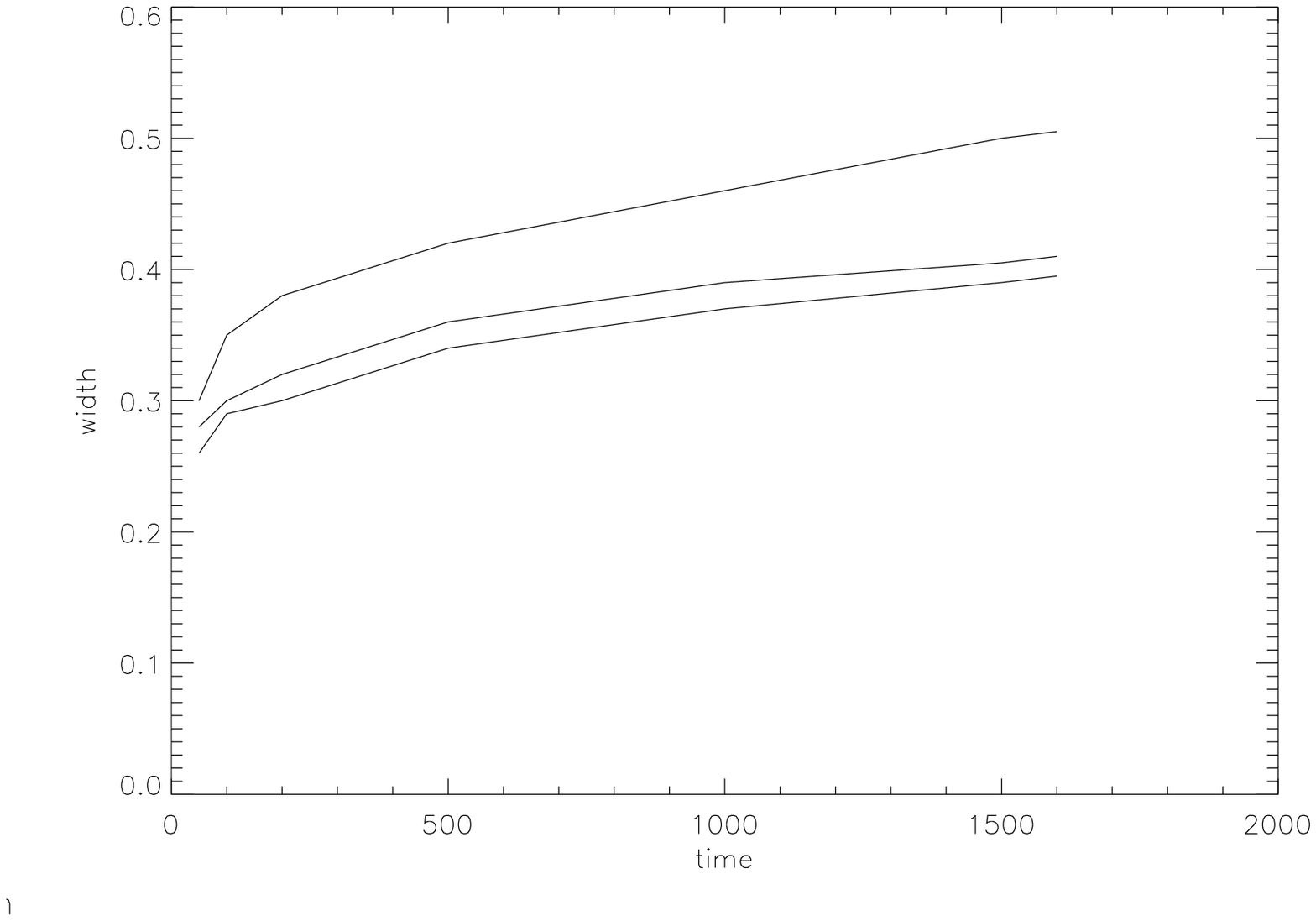}
\caption[]{
\baselineskip14pt
Density profiles and width measurement.
a) Disk radial density profiles are shown at different times
for a simulated protoplanet of mass ratio
$q=10^{-3}$ with a Reynolds number of $2\times 10^{6}$.
Profiles are shown for times $t=100,200$ and 1000 in units
of planet's orbital period.
The density profiles are shown with 3 different dotted lines which are
used to define our measurements of the width between
the planet and the disk edges.
The upper line is the initial density profile, the intermediate one is half
of the initial density and the lowest one
is set at a density of $10^{-6}$, about one order of magnitude lower
than the initial density profile.   Our first type of measurement
uses the location at which the density profile crosses the first line.
Likewise for the second and third types of measurements.
We measure a width between the location of the planet
and the location where these lines intersect the density profile.
b) The width between the planet and the gap outer boundary computed with the three
procedures defined in a) are shown as a function of time.
We see that the two bottom lines exhibit similar behavior.  The upper
line is sensitive to the shape of ``top'' of the outer disk edge.
Because of this sensitivity we use the third type of measurement (lower line)
in our subsequent figures
to quantity the distance $\Delta$ between the planet and outer disk edge.
\label{fig:gap_def}}
\end{figure*}

To measure the distance between the planet and the edge
of the disk from a numerical simulation we must first
consider procedures for defining a point of measurement
for the disk edge.
As seen in Figure \ref{fig:gap_def}a,
the gaps are not well approximated by a square well.
Figure \ref{fig:gap_def}a illustrates three procedures for
defining a location to measure the disk edge from a radial density
profile. In each case we choose a stationary radial function and
define $\Delta$ based on the location where the radial density
profile crosses this function.  Our first case, defined by the
upper line in Figure \ref{fig:gap_def}, is the initial unperturbed
density profile.  For the second case we use one half of the
initial density profile to define our points of measurement. Our
third case (represented by the horizontal line in Figure
\ref{fig:gap_def}) is defined with an absolute density of
$\Sigma = 10^{-6}$ (independent of radius) which is approximately
an order of magnitude below the initial disk density at $r=r_p$.

We now compare the evolution of $\Delta$ based on these
measurements. From Figure \ref{fig:gap_def}b we see that the
second and third definitions evolve similarly, but the first one,
which refers to the initial density profile, continues to
increases on long timescales to a greater extent than the other
measurements. Because accretion continues from the outer parts of
the disk, the gas density at the outer boundary of the gap continues
to increase on long timescales. It is likely that this slow
accretion governs the longterm behavior at the ``top'' of the
density profile. To minimize the sensitivity to the exact shape of
the edge profile we subsequently choose to measure the distance
between the planet and the edge of the disk using our third
measurement case (referring to an absolute density of $\Sigma =
10^{-6}$). Because there are no significant differences between the
second and third measurements, we have chosen to subsequently
measure the simpler of the two.

\subsection{Evolution of the disk edge}

\begin{figure*}
\epsscale{0.60}
\plotone{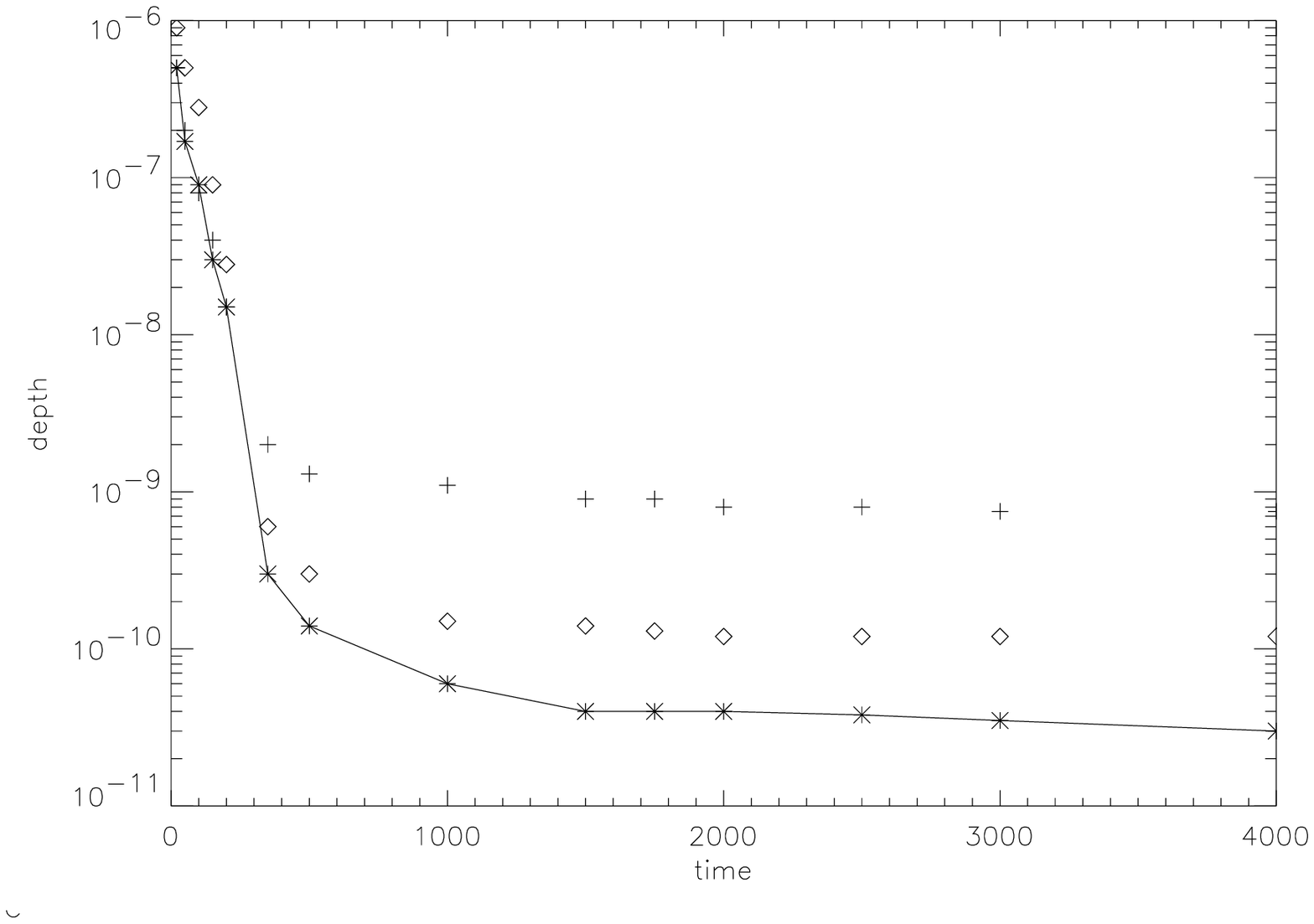}
\plotone{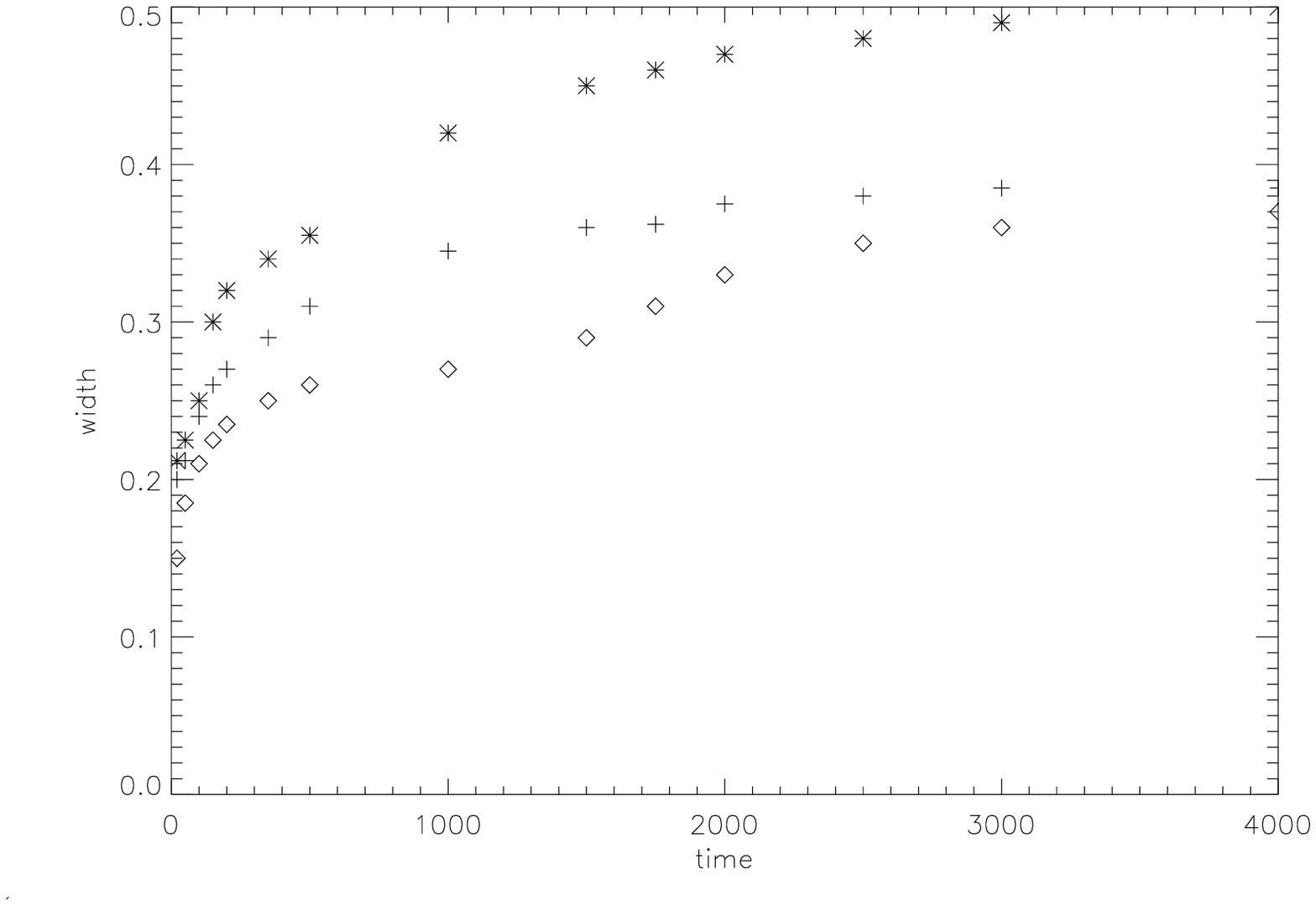}
\caption[]{
Gap depth and width as a function of time.
a) The log of the disk density near the planet as a function of time ($x$-axis)
for a planet mass ratio of $q = 2 \times 10^{-3}$ (stars, connected), and
$q = 10^{-3}$ (crosses).  For these simulations the Reynolds number is
${\cal R} = 2\times 10^6$.
Diamonds are shown for a simulation at higher viscosity simulation
with $q = 2 \times 10^{-3}$ and ${\cal R} = 2 \times 10^5$.
Time is given in units of the planet's orbital rotation period.
The density is given in units of the star's mass divided by the square of the
semi-major axis of the planet.
b) Distance of planet to the edge of the disk (defined as that having
a density of $10^{-6}$) as a function of time,
for $q = 2 \times 10^{-3}, {\cal R} = 2\times 10^6$
(starred points), $q=10^{-3}, {\cal R} = 2\times 10^6$  (crosses)
and $q=2 \times 10^{-3}, {\cal R} = 2 \times 10^5$ (diamonds).
The width is given in units of the semi-major axis of the planet.
\label{denwin_time}}
\end{figure*}

Figure \ref{denwin_time} shows the evolution of $\Delta$, the distance
between the planet and the disk edge, and the
density at the location of the planet for three simulations (runs
\#6, \#8 and \#11 listed Table 1) which share either the same
planet mass ratio or viscosity.  The starred points  show the
depth and width for a planet mass ratio of $q = 2 \times 10^{-3}$
in a disk with ${\cal R} = 2\times 10^{6}$.  We refer to this run
as the ``reference run''. The crosses show a run that has the
same Reynolds number but a planet mass ratio $q = 10^{-3}$, half
that of the reference run. The last simulation (shown with diamond
points) has the same planet mass ratio as the the reference run
but a higher disk viscosity (${\cal R} = 2\times 10^{5}$).

Our initial density distribution, that of a smooth power law, is
not an equilibrium state in the presence of a planet. As previous
numerical studies have illustrated, the planet accretes gas and
tidally induces spiral density waves that repel the gas from the
planet. Figure \ref{denwin_time} shows that two regimes govern the
evolution of the depth and width of the gap.  On short timescales
($t< 200$ orbital periods) the clearing of gas near the planet is
rapid. The depth of the gap is almost independent of the planet
mass and viscosity, however the distance between the planet and
the gap outer boundary of the disk is dependent upon both.    After $t
\sim 200$ orbital periods, both the density and gap width evolve
much more slowly, as if approaching a steady state.

A careful inspection of Figure \ref{denwin_time} shows that the initial
evolution of the simulations
are very similar, and nearly independent of the disk viscosity.
This suggests that at early times the angular momentum flux is dominated by
spiral density waves driven by the planet, and that the viscous inflow rate
is relatively minor.
The clearing of the initial gap should
take place on a timescale that is set by how fast the spiral density
waves can push mass away from the planet.
The torque on the disk can be estimated from
the sum over each resonance outside the edge of the disk.
If the density drops near the planet, then the Lindblad resonances
nearest the planet will not significantly contribute to the torque
on the disk. In other words, because equation (\ref{T_m}) depends on the density
at the resonance, only resonances outside the gap in the disk will
exert a significant torque on the disk.

At the beginning of the simulation the gas disk extends all the way
to the Roche lobe of the planet and the clearing timescale is set
by a summation over high order (large $m$) Lindblad resonances.
However as gas is cleared away from the planet, the density
drops at the highest order resonances and the lower order Lindblad resonances
take over. Since these aren't as strong, this clearing should go
at a slower rate.  So we expect that much of the clearing takes place
under the influence of only the last few resonances.
From equation(\ref{Tmfm})
we expect evolution on a timescale in planetary rotation periods
\begin{equation}
t_{clear} \sim
{r_p^4 \Omega_p^2 \Sigma \over 2 \pi \sum_m T_m}
      \sim 250 \left({q \over 10^{-3}}\right)^{-2}
\end{equation}
where we have added the contributions from the $m=$2,3 and 4
Lindblad resonances.

When the gap opens
sufficiently that the higher order resonances are evacuated and
the $m=2$ OLR resonance dominates, the timescale could be an order of
magnitude longer. Nevertheless, the above timescale is
approximately what we see for the first phase of evolution from
the simulations presented in Figure \ref{denwin_time}, and is
consistent with this first phase of evolution being independent of
the disk viscosity. Since the timescale for clearing the gas away
from the planet depends primarily on the planet mass, we expect
that the simulations with massive planets will initially clear
material faster than those with lower mass planets.  This is
primarily seen in the evolution of the disk edge (Figure
\ref{denwin_time}b) and not in the depth of the gap (Figure
\ref{denwin_time}a), though the depth of the gap in the lower
planet mass simulation (crosses) is higher than that of the higher
planet mass simulation at the same viscosity (stars).

On longer timescales viscous behavior becomes important. For the
higher viscosity simulations we expect the viscosity becomes
important earlier on, consistent with the quicker transition
toward the regime of slower evolution seen in the higher viscosity
simulation (diamonds) in both density and width plots of Figure
\ref{denwin_time}. Except for the lowest viscosity simulations
considered in this paper, the transition to the longer slower
evolution takes place near $t=200$ orbital periods. So
measurements taken after this time can be considered to represent
quantities measured when the disk is slowly evolving. Because
accretion continues from the outer disk, gas continues to pile up
at the edge of the disk.
Consequently these simulations never reach a steady state, but
on long timescales slowly evolve on the viscous timescale
of the outer disk.

The long term accretion complicates the development of an
understanding  of the scaling
behavior of the edge profile on planet mass and disk viscosity. To
enable us to explicate the dependence of the disk edge profile on
planet mass and disk viscosity we will subsequently compare gap
properties measured at different times and discuss the scaling of
the properties of disk edges keeping in mind that our simulations
show evolution on two timescales.

\subsection{Influence of the disk's viscosity}

Figure \ref{fig:Reynolds} shows the depth and distance to the edge of the disk
for a simulation with
planet mass ratio $q= 2 \times 10^{-3}$ as function of the Reynolds number.
The upper curve (diamond points) is at $t=200$, {\it i.e.} in the first phase of
evolution. The lower curve (starred points) is at $t=1000$, and shows the
second phase of evolution.
At latter times when the viscous timescale becomes important,
we see that the depth is more strongly influenced by the
disk viscosity than during the first phase.
The depth at $t=200$ is independent of viscosity
for ${\cal R} \gtrsim 10^6$ because the secondary timescale evolution has
not yet completely taken over (see Figure \ref{denwin_time}).

Although the gas density near the planet is strongly dependent
on the viscosity at later times, the distance
between the planet and the edge of the disk is
only weakly dependent upon the disk viscosity
(see Figure \ref{fig:Reynolds}b).
A variation of 2 orders of magnitude
in the disk viscosity causes only a
30\% change in the distance between the planet and disk edge,
but a similar scale (2 orders of magnitude) variation in the gas density.
The strong dependence of the gas density near the planet on viscosity,
and weak dependence of the actual location of the disk edge on
the disk viscosity will be discussed in section 6 when we consider
modifications to the previous theory.

\subsection{Influence of the protoplanet's mass}

Figure \ref{fig:q} shows the behavior of the gap and disk edge
created in a disk
of viscosity corresponding to ${\cal R} = 2\times 10^{6}$
as function of the planet
mass ratio.  We plot results for simulations carried out
with planet mass ratios between $q = 10^{-4}$ and
$2 \times 10^{-3}$.
The two lines shown in both figure panels are similar so
we can study the dependence of the gap density and width
on the planet mass ratio without being sensitive to the
time at which the measurements were taken.

We see from Figure \ref{fig:q}
that both the depth and the distance between the planet
and disk edge vary
significantly as function of the planet mass even though we have
only run simulations over a relatively small range in planet mass
ratio (particularly when we compare to the larger range of
Reynolds numbers that we explored).
Since a gap will not open at the lowest planet masses and higher ones
exceed the masses of most known extrasolar planets, we cannot
significantly extend the range of planet masses considered.
In the regime plotted in Figure \ref{fig:q}
we see that a variation in the planet mass ratio of a factor of 4 causes
approximately an order of magnitude a change in the gas density
near the planet, and a 40-50\% change in the distance between
the planet and the disk edge.
As we previously found for the dependence on disk viscosity,
the distance between the planet and the disk edge is less
strongly dependent upon the planet mass ratio than
is the gas density near the planet.
Both the depth and distance between
the planet and disk edge are more strongly dependent upon the planet
mass ratio than on the disk Reynolds number.
In the following section we will attempt to account for
the trends we observe in the long timescale evolution
of these simulations.

\section{Comparisons to Theory }

\begin{figure*}
\vspace{8.0cm}
\vskip 10cm
\includegraphics{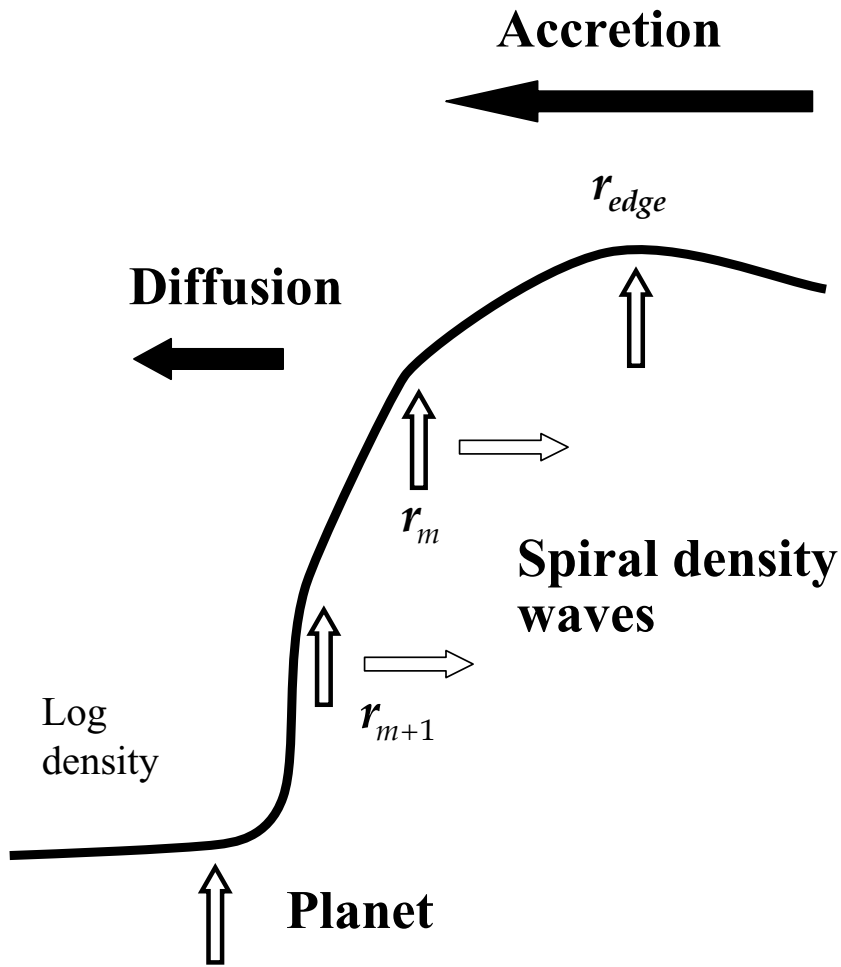}
\caption[]{
We expect that diffusion or viscous spreading is important
in the steep edge of the disk.
Arrows represent the direction of the mass flux.
The torque from dissipated spiral density waves is balanced by
by diffusion in the steep edge of the disk.   To maintain a balance,
the density at the resonance can be very low.
In the regime
where the distances between the resonances is smaller than the dissipation
scale length, each resonance cases a change in the slope of the density
profile.
\label{cartoon}}
\end{figure*}

As suggested by \citet{bryden}, when mass has been removed from
the region nearest the planet and a steep density gradient is
present, diffusion or viscous spreading may become important and
determine the shape of the edge of the disk (see Figure
\ref{cartoon}).
The basic equation that describes the evolution of a Keplerian
disk due to viscous processes is
\begin{equation}
{\partial \Sigma \over \partial t} = {3 \over r}  {\partial \over
\partial r}
     \left[ r^{1/2} { \partial \over \partial r} \left( \nu \Sigma r^{1/2} \right)
     \right]
\label{pringle}
\end{equation}
\citep{pringle81}.  This equation arises by considering the
Navier-Stokes equation and conservation of mass.  The azimuthal
component of the Navier-Stokes equation, which addresses angular
momentum transport, is used to estimate the radial velocity
component. This, when inserted into the equation for conservation
of mass, yields the above equation. When the density $\Sigma$ is
nearly flat, the angular momentum flux or torque through a radius
$r$ is given by equation(\ref{T_nu}).  This is what is usually
meant by ``accretion,'' as we shall see.  When the radial gradient
of $\Sigma$ is very large however, there is an additional
diffusive term that drives mass flow.

The previous equation can be modified to include mass and angular
momentum flow due to dissipation of spiral density waves driven
by the planet by adding a term,
\begin{equation}
{\partial \Sigma \over \partial t} = {3 \over r}   { \partial
\over \partial r}
     \left[ r^{1/2} { \partial \over \partial r} \left( \nu \Sigma r^{1/2} \right)
               + { 1\over 3 \pi r \Omega}\sum_m { \partial F_m \over \partial r}
     \right]
\label{disk_evolution}
\end{equation}
where ${ \partial F_m \over \partial r}$ corresponds to the torque density
exerted in the disk due to dissipation of spiral
density waves driven by the planet
at the $m$th Lindblad resonance, and there is
a sum of terms due to the contributing resonances. Here we
have followed the derivation by \citet{takeuchi}.

Once a prominent gap is opened we must distinguish between the
physics occurring within the gap, (particularly along its steep
outer edge), and that occurring outside the gap edge at larger
radii. Outside of the gap the density gradient is shallow, the
diffusion rate will be small and we expect that the mass inflow
due to accretion will dominate and will be balanced by mass
outflow resulting from the spiral density waves. However, in the
steep edge itself, diffusion should be important. In this region
it is mass inflow due to diffusion that should be balanced by the
mass outflow caused by dissipation of the spiral density waves.
Thus we can divide the
problem into two regimes: accretion dominated and
diffusion dominated (see Figure \ref{cartoon}).
This division can be directly seen in Figure \ref{fig:gap_def}a showing
the density profile of a single simulation.
Past the edge of the disk ($r\sim 1.3$)
the density gradient is small, but inside this radius, the density drops
by orders of magnitude and the density gradient is large.

Unless the planet mass is sufficiently low and viscosity
sufficiently high that a gap is only barely opened, we expect that
edge of the disk will be steep.
The resonance most distant from the planet at highest density $\Sigma$ should
dominate the shape of the outer part of the edge
because it will be most effective at driving spiral density waves.
We refer to quantities such as radius and
density past the edge of the disk as $r_e$ and $\Sigma_e$ and
those at the $m$-th OLR resonance as $r_m$ and $\Sigma_m$
(see Figure \ref{cartoon}).

In the limit that the density is not varying quickly with time,
(${\partial \Sigma \over \partial t} \approx 0$),
the previous equation can be written
\begin{equation}
{3 \nu } \Sigma +  3 \nu r { \partial \Sigma \over \partial r}
 \sim
-{1 \over  3 \pi r \Omega}
\sum_m {\partial F_m \over \partial r  }    + {\rm constant}
\label{disk_steady}
\end{equation}
where the first term on the left is what is typically known as
accretion\footnote{
The first term is $3 \nu r \Sigma/2$ when $\nu$ is independent of $r$
leading to equation \ref{T_nu}, however here we have assumed
that the Reynolds number is constant, consistent with an "$\alpha$-disk".}
and the second can be described as diffusion.
Here $F_m$ is the angular momentum flux transfered by the waves
and $-{\partial F_m \over \partial r  }$ is the torque density exerted
on the disk due to the dissipation of the waves in the disk \citep{takeuchi}.
Using the definition for the Reynolds number, we can rewrite the previous
equation as
\begin{equation}
{\partial (r \Sigma)\over  \partial r}
\sim -{{\cal R} \over 3 \pi G M_*} \sum_m {\partial F_m \over \partial R}.
\end{equation}
Integrating the previous equation we find
\begin{equation}
 r \Sigma \sim -  \sum_m {{\cal R} F_m \over 3 \pi G M_*}  + {\rm constant}
\end{equation}
which implies that the shape of the disk edge is entirely determined
by the way that the waves are dissipated.  The steep edges we
see in the simulations imply that the dissipation occurs very rapidly,
as previously pointed out by \citet{takeuchi}.

The angular momentum flux transferred by the spiral density waves
$F_m$ is equal to that given by equation (\ref{T_m}) at the
location of the resonance $r_m$ and then decays because of
dissipation. We assume that the spiral density waves are
dissipated over a length scale $r_d$ (referred to as a damping
length by \citealt{takeuchi}), so $F_m \sim T_m
(1-e^{-(r-r_m)/r_d})$ for $r> r_m$ and $T_m$ is given by equation
(\ref{Tmfm}) with quantities defined at the resonance. We
associate the edge of the disk as that point where the dissipation
of the density waves ceases to be important; $r_e \sim r_m + r_d$
where $r_m$, in this case, is the outermost resonance that still
lies within the disk edge.

Considering only one resonance we can rewrite the previous equation as
\begin{equation}
 r \Sigma \sim  r_m \Sigma_m
+  {{\cal R} \Sigma_m f_m r_p \over  3 \pi}
  \left({q \over 10^{-3}}  \right)^2
\left(1 - e^{-(r-r_m)/r_d}\right)
\label{sum_steady}
\end{equation}
for $r>r_m$.
We expand this to include all resonances.
\begin{equation}
 r \Sigma \sim  r_p \Sigma_p  + \sum_m
  {{\cal R} \Sigma_m f_m r_p \over  3 \pi}
  \left({q \over 10^{-3}}  \right)^2
  a_m(r)
\label{all_resonances}
\end{equation}
where
\begin{eqnarray}
a_m(r) =
& 0                      & {\rm for} ~~~   r<r_m  \nonumber \\
& 1 - e^{-(r-r_m)/r_d}   & {\rm for} ~~~   r>r_m   .
\end{eqnarray}

First let us consider the outermost resonance that still
lies within the edge of the disk.
Since the rightmost term of this equation is large for $r > r_m + r_d$,
we can estimate the density ratio at the edge compared
to that at the resonance
\begin{equation}
{\Sigma_e \over \Sigma_m} \sim {{\cal R} f_m \over  3 \pi}
  \left({q \over 10^{-3}}  \right)^2.
\label{density_ratio}
\end{equation}
We find that the drop in density near the disk edge is dependent
on the square of the planet mass ratio and the Reynolds number.
This is consistent with our numerical exploration which found that
the gap depths were strongly dependent on both quantities  and
more strongly dependent on the planet mass ratio than the disk
viscosity (section 4.3, 4.4). However, the distance from the
outermost resonance to the peak density, $\Sigma_e$, only depends
on the scale over which the dissipation takes place, $r_d$. Using
the WKB approximation, \citet{takeuchi} found that the dissipation
scale length was primarily dependent on the Mach number of the
disk, but also dependent on the disk viscosity. The weak
dependence in our simulations we found for the distance of planet
to the disk edge implies that the dissipation length is not
strongly dependent on the viscosity.  This may in part be
due to the formation of shocks (e.g., \citealt{ward97}).

\begin{figure*}
\epsscale{0.60}
\plotone{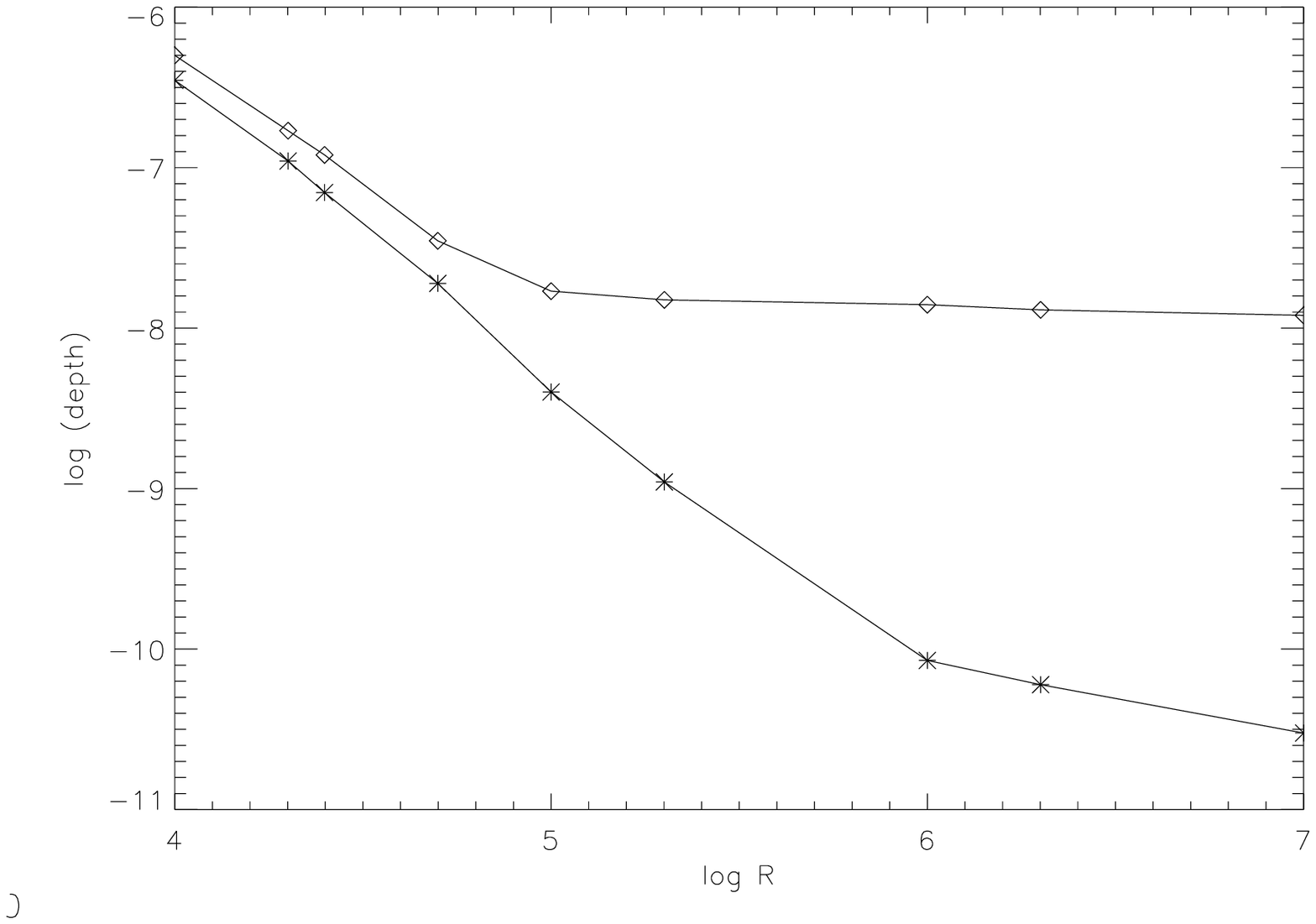}
\plotone{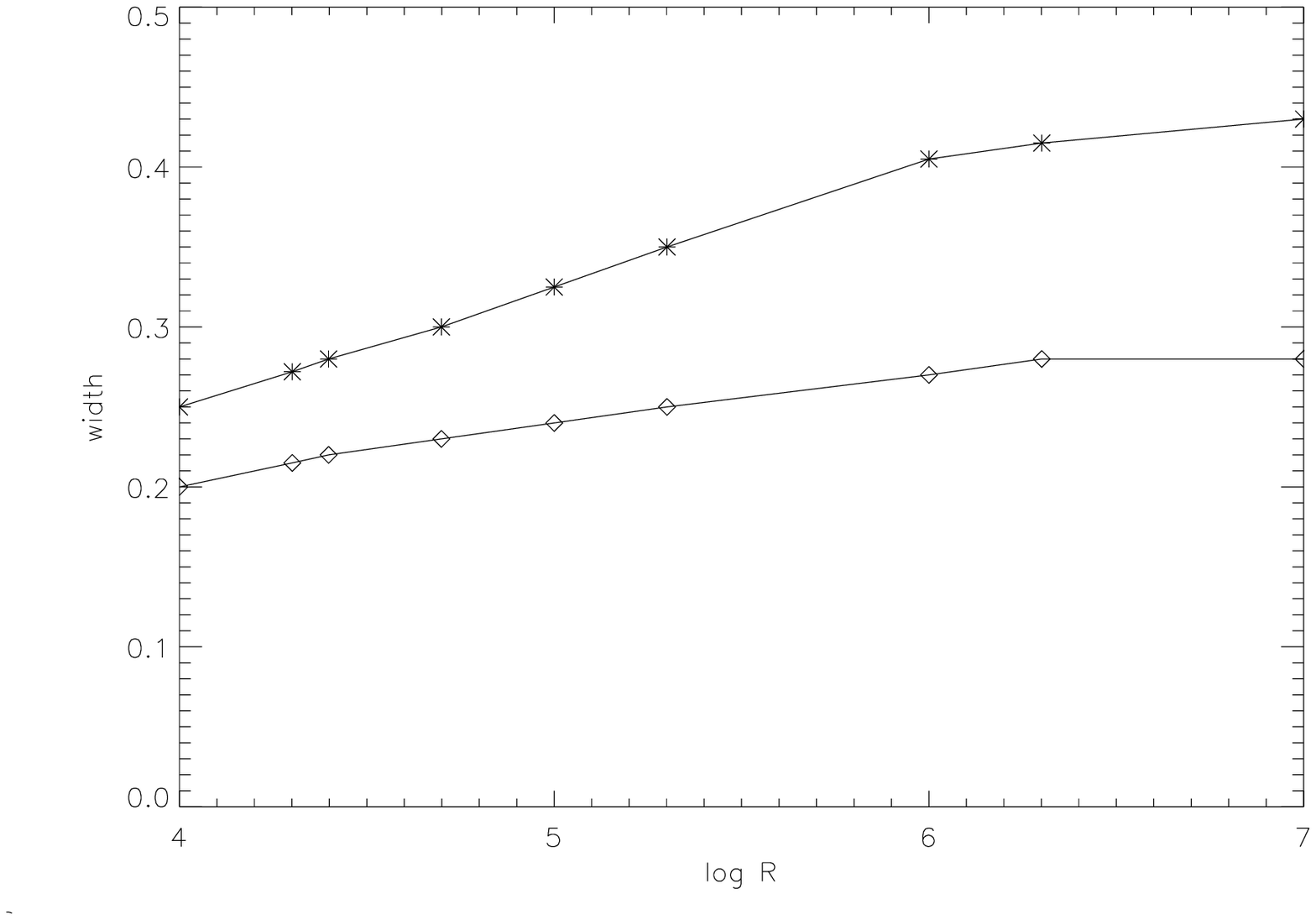}
\caption[]{ Gap properties as a function of disk viscosity.
a) Depth of the gap  as a function
of Reynolds number at two times $t=200$ (diamonds) and $1000$ (stars)
orbital periods
for simulations with planet mass ratio $q=2\times 10^{-3}$.
For comparison the initial density near the planet was
$\Sigma \sim 2 \times 10^{-6}$.
b) Distance from planet to the edge of the disk
(defined as that having a density of $10^{-6}$) as a function
of Reynolds number at the same two times $t=200$ and $1000$.
We find that the depth is a strong function of Reynolds number, but
that the distance between the planet and edge of the disk is not
a strong function of Reynolds number.
\label{fig:Reynolds}}
\end{figure*}

\begin{figure*}
\epsscale{0.70}
\plotone{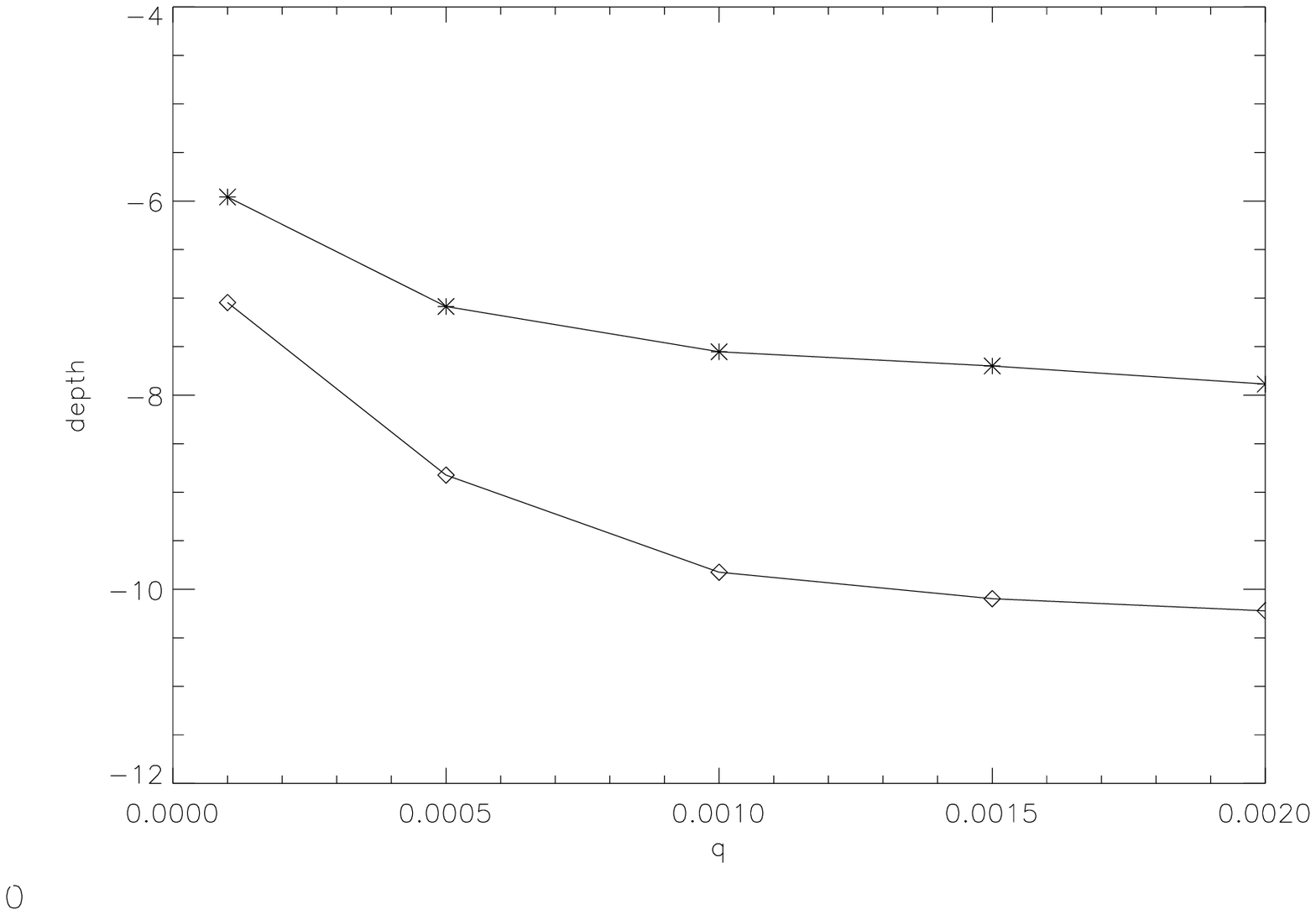}
\plotone{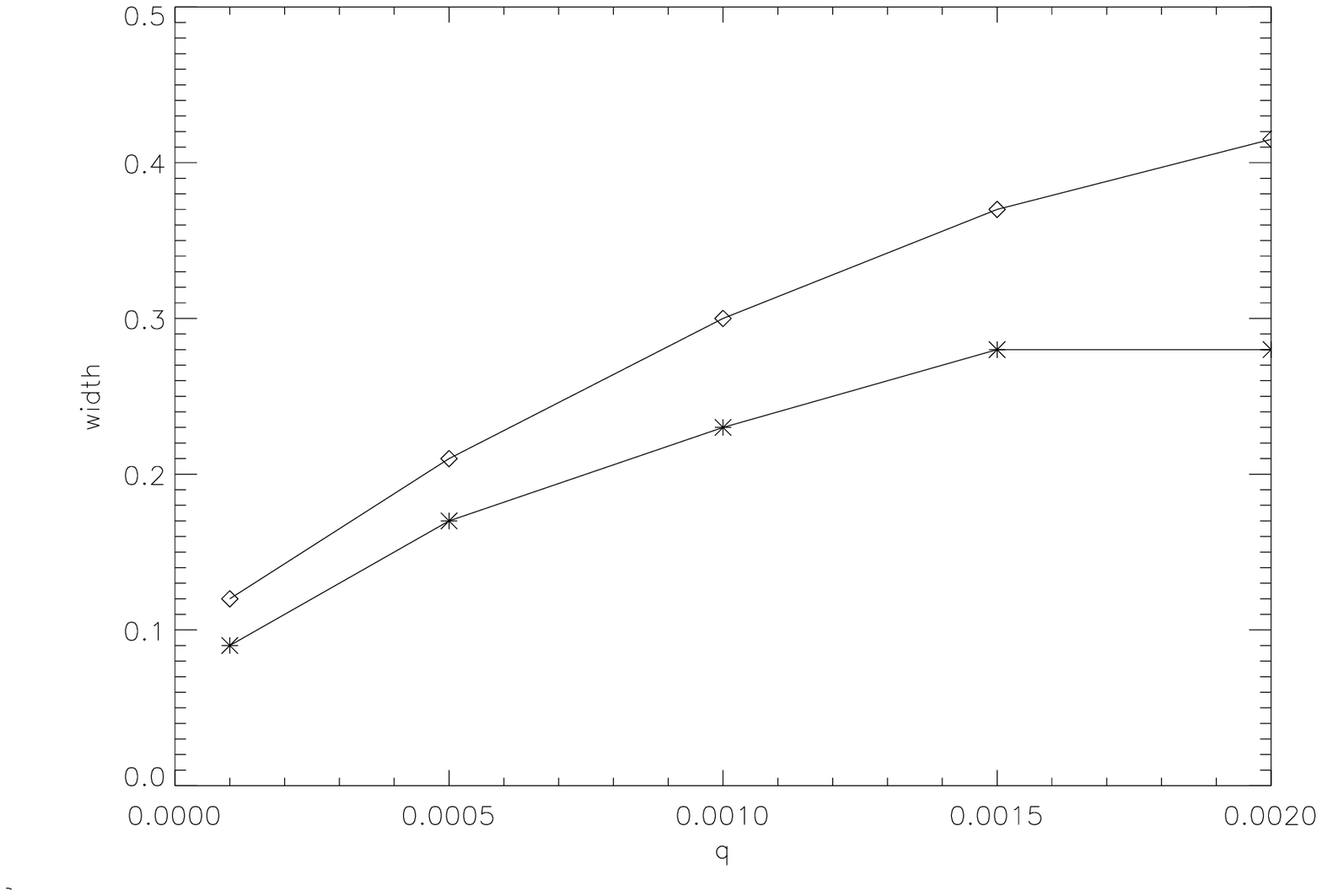}
\caption[]{ Gap properties as a function of planet mass.
a) Depth of the gap  as a function
of planet mass ratio, $q$,  at times $t=200$(stars) and $1000$ (diamonds)
orbital periods, for simulations
with Reynolds number ${\cal R} = 2\times 10^6$.
b) Distance from the planet to the edge of the disk
(defined as that having a density of $10^{-6}$)
as a function of the mass ratio at times $t=200$ and $1000$.
We find that the density near the planet is a strong function of planet
mass ratio, however the distance between the planet and edge of the disk
is not a strong function of planet mass ratio.
\label{fig:q}}
\end{figure*}

Equation (\ref{density_ratio}) allows us to estimate
the density at the radius of the dominant
Lindblad resonance from the following parameters; the resonance
parameter $f_m$, the planet mass ratio, $q$, and the Reynolds number, $\cal R$.
We consider the simulation displayed in Figures
(\ref{fig:Reynolds}, \ref{fig:q}) with planet mass ratio
$q= 2\times 10^{-3}$ and ${\cal R} = 2 \times 10^6$.
Because the gap outer edge
is nearest the $m=2$ OLR, we expect that this resonance, with
$f_2 = 5\times 10^{-5}$, balances the viscous inflow.
From  these parameters and using equation (\ref{density_ratio})
we estimate ${\Sigma_e \over \Sigma_2} \sim 42$.
This ratio and the scale length $r_d$ determines the slope in the outermost
brightest and  most easily detectable region of the disk edge.
While the ratio $\Sigma_e \over \Sigma_2$ is greater than 1, it is
not  high enough density ratio to explain the orders of magnitude
lower densities seen the planet that we measured in the
simulations (see Figures \ref{fig:Reynolds},\ref{fig:q}).
Consequently we conjecture that more than one resonance must
affect the the disk edge profile.

To consider the role of the different resonances we look more
closely at the density profile shown in figure \ref{fig:gap_def}a.
Equation (\ref{density_ratio}) allows us to estimate the change in
density between the outermost resonances. For $q=10^{-3}$, ${\cal
R} = 2 \times 10^6$, we expect ${\Sigma_e \over \Sigma_2} \sim
11$, accounting for order of magnitude drop in density between
$r=1.3$ (the location of the $m=2$ OLR) and $r\sim 1.5$ (the
location of peak density) as is consistant with what is seen in
figure \ref{fig:gap_def}a. Between the $m=3$ OLR (at $r=1.2$) and
the $m=2$ OLR (at $r=1.3$) we expect the density to change by a
factor of ${{\cal R} f_3 \over 3 \pi} \left({q \over 10^{-3}}
\right)^2 = 42$ which again is consistant with for the drop in
density between $r=1.3$ and $r=1.2$ of another order of magnitude.

Thus it appears we can divide the disk edge into two regimes, one
it which the distance between the resonances is larger than the
dissipation scale length $r_d$ and another where the distance
between the resonances is smaller than $r_d$.  Note that the
distance between the resonances increases as a function of
distance from the planet and is inversely dependent on $m$. When
the distance between the resonances is larger than the dissipation
scale length then the waves can fully dissipate their angular
momentum and the density increases by a factor of ${{\cal R} f_m
\over  3 \pi}
  \left({q \over 10^{-3}}  \right)^2$.
Each resonance causes an increase in density by a multiplicative
factor which depends on the Reynolds number, the square of the
planet mass and $f_m$.

We now consider the regime nearer the planet where the distance
between the resonances is smaller than the dissipation scale
length.  In this regime we the waves do fully dissipate their
angular momentum before the next resonace is encountered.  We
conjecture that the effect of the resonances is additive in this case
rather than multiplicative. Note here that when the resonance Mach
number ${r \Omega \over m c} < 1$ the resonances can no longer
efficiently drive spiral density waves. Thus there is a cutoff in
$m$ and a lower limit on the density near the planet.

Because the higher order resonances only additively affect
the density profile, the ratio of the density near the planet
compared to that outside the disk edge is primarily
determined by the number of resonances in the outer regime,
where the distance between the resonances is smaller than the
dissipation scale length and the density increases
by a factor of ${{\cal R} f_m \over  3 \pi}
  \left({q \over 10^{-3}}  \right)^2$ between each resonance.
The high power dependence of the ratio of the density  between the
planet and disk edge on $q$ and on ${\cal R}$
(that we remarked on in sections 4.3 and 4.4;
see Figures \ref{fig:Reynolds} and \ref{fig:q}) can be explained
if more than one resonance is in this regime, and if $r_d$ is
not strongly dependent on the planet mass or disk Reynolds number.

In contrast, we expect the location of the disk edge to be set
by the outermost resonance capable of driving significant density
waves and the dissipation scale length $r_d$.
From equation (\ref{all_resonances}) we expect that
the outermost resonance
is the lowest $m$ for which  ${{\cal R} f_m \over  3 \pi}
  \left({q \over 10^{-3}}  \right)^2 \gtrsim 1$.  This limit
is approximately consistent with the distance between the planet
and disk edge dropping below 0.3 for $q \lesssim 10^{-3}$ seen in figure
\ref{fig:q}b.
In section 4.3 and 4.4 we found that the distance between the planet
and disk edge was only weakly dependent on the planet mass and
disk viscosity.
Once the outmost resonance is set, equation(\ref{all_resonances}) suggests that
that this distance only depends on the
dissipation scale length.  If this is not
strongly dependent on the disk visocity then we can account
for weak dependence of this distance on planet mass and Reynolds number.

\section{Summary and Discussion}

In this paper we have presented a series of 2D numerical
hydrodynamical simulations of planets embedded in Keplerian disks.
We have covered a different area of parameter space compared to
previous works, and have concentrated on lower viscosities or
higher Reynolds number disks with the expectation that this regime
would represent a more extended time after planets have formed and
have ceased to accrete gas from the protostellar disk.  As
previous simulations have shown, the planet can open a gap in the
disk. On longer timescales we find the width of the gap depends
only weakly on the mass of the planet and disk viscosity. The
depth of the gap and slope of the outside edge of the gap is,
however, comparatively more strongly dependent on both quantities,
and a stronger function of planet mass ratio than of viscosity.
When $q^2 {\cal R }\gtrsim 0.3$, the edge of the disk can be found
between the $m=2$ OLR and the $m=1$ OLR.  This is unexpected since
the planet can only resonantly drive spiral density waves at the
locations of resonances.

We account for the phenomena seen in the simulations with a scenario
model  which balances the viscous transport of
angular momentum against that arising from the dissipation
of spiral density waves induced tidally by the planet.
The disk approaches a steady state where outside the gap
the inward mass flux from viscous accretion is balanced by
outward flux due to the dissipation of
planet driven spiral density waves.  Within
the steep gap edge, however, it is diffusion that balances the
outward angular momentum flux from the dissipation of the spiral density
waves. These waves are excited inefficiently because of the
extremely low gas density near the planet.
The profile shape is determined by the scale length
over which the spiral density waves are dissipated, $r_d$.

From our simplified model we conjecture that the edge can be
divided into two regimes: where the distance between resonances is
greater than $r_d$ and where this distance is smaller than the
dissipation scale length. In the outermost regime each resonance
multiplicatively increases the density as a function of distance
from the planet. The multiplicative factor depends on the product
of the Reynolds number and square of the planet mass.  This
prediction matches the stronger dependence of gap properties on
planet mass ratio then on Reynolds number that we also observed in
the simulations.  Finally the distance between the planet and the
disk edge is set by the outermost resonance capable of driving
strong density waves; the $m$-th OLR for which ${{\cal R} f_m
\over  3 \pi}
  \left({q \over 10^{-3}}  \right)^2 \gtrsim 1$ and by the dissipation
scale length.  The weak dependence we saw for this distance on the
planet mass and Reynolds number suggests that the dissipation
scale length is not strongly dependent on the disk viscosity. Once
the outermost resonance is determined, the slope of outer, most
easily detected part of the disk edge depends on $r_d$ and the
ratio ${{\cal R} f_m \over  3 \pi}
  \left({q \over 10^{-3}}  \right)^2$.

\citet{rice} noted that the slope of a disk edge maintained
by a planet is dependent upon the planet mass
and suggested that this dependence could be used to constrain the
type of planet that could be present from fits to the spectral
energy distribution (as from GM Aurigae). The scaling
we have developed confirms this work and furthermore suggests that
the slopes of disks edges maintained by planets are approximately
proportional to the product of the planet mass ratio squared and
the Reynolds number.

The large drop in density in the disk edges predicted by our
simulations implies that the opacity of the gas near the planet is
much reduced.   Because disk edges are illuminated by the central
star, and could be optically thin, the spectral energy
distribution predicted from the simulations may be sensitive to
the slope of the edge. As proposed by \citet{rice}, its possible
that circumstellar disk edges may be used to constrain the masses
of planets which maintain them. We note that because the outer
disk slope depends on both $r_d$ and a density ratio which is set
by the outermost resonance, there could be more than planet mass
setting observed outer edge slopes. Future work could focus on how
the scale length $r_d$ affects the slope and so might discriminate
between possible degenerate solutions. One interesting possibility
is that the changes in gas density slope can be detected. The
ratio of the slopes on either side of the transition point is set
by the resonance at the transition region and the distance from
the transition point to the density peak is set by the dissipation
scale length. If  such measurements became possible then the
degeneracies could be lifted and the planet mass tightly
constrained from the disk edge density profile.

Our simulation results clearly point to the need for a deeper
understanding of the physics of gaps and disk edges.
Simplistic models such as explored in section 2,
that neglect diffusion and the density gradient in the torque balance
estimate, would predict a stronger scaling on the gap size with $q$ and $R$
than we have measured from our simulations.
The diffusive model we discuss here is preliminary
because it is used to approximate long timescale behavior seen in
the simulations which never actually achieve a steady state. We have
neglected variations in the angular rotation rate caused by the
pressure differential in the steep edge of the disk.
We estimate
however that the location of the resonances are not significantly
affected by this pressure differential. Our model also assumes
that spiral density waves are driven at the location of the
resonance and we have not considered models that drive density
waves at significant distances from Lindblad resonances (e.g., as
exhibited in the simulations of \citealt{savonije}).  The ability
of a planet to drive density waves off resonance depends on the
sound speed of the disk (and possibly on the boundary conditions)
and it is more likely this would occur in
low Mach number disks. In this situation we expect that the gas
density near the resonances would be reduced to even a larger
extent to balance viscous inflow in the high planet mass, low
viscosity simulations. The importance of this driving mechanism
and the resulting affect on the disk edge
can be explored by carrying out simulations of thicker disks.
By affecting the radial range over which the spiral density waves
are launched, the thickness of the disk may influence
the shape of the edge profile.  The simulations
of \citet{kley} suggest that thicker disks have shallower
disk edges. This sensitivity, which we have not yet explored,
would increase the number of parameters required to describe
a disk edge.  Here we have adopted
an exponential form for the dissipation of the waves with radius.
However the sensitivity of the dissipation scale length to
sound speed and viscosity has not yet been explored  beyond
that discussed by \citet{takeuchi} and the sensitivity of
the disk edge profile to these parameters could be more carefully explored.
Our work has been restricted to the case of a planet in a circular
orbit, however the interaction between a planet and a disk
may induce eccentricity in the planet \citep{sari}.
In this case,
because of the increased number of resonances at which spiral density
waves can be driven, it should be even more difficult to predict
the shape of the disk edge profile.

Because mass continues to pile up outside the outer
edge of the disk near a planet, a steady state will not be reached, increasing
the difficult of predicting the edge profile.
The density near the planet will increase until the planet begins to
migrate. As disks age and the viscosity drops, the timescale for
planetary migration times will also increase, so observed
protoplanetary and debris disks may be slowly evolving rather than
steady state.  Our partial understanding based on the long
timescale behavior of these systems gained from this study may be
relevant toward the interpretation of disk edges observed in
forthcoming observations. Since the edges of disks receive and
process radiation differently than disk
surfaces, observations may be particularly sensitive to the
3 dimensional shapes of these edges.  Future work should also
concentrate on developing 3D models for disk edges in the physical
regime studied here.

Finally, in a model where planetary formation is sequential, it is
tempting to speculate that the gas that piles up on long
timescales outside the $m=2$ OLR may induce
formation of an additional planet (\citealt{pp3}).
Our study suggests that the secondary planet would seldom form
locked in the 3:2 mean motion resonance with the primary as is
Pluto with Neptune (at $r\approx 1.3$,
the 3:2 mean motion resonance is equivalent to
the $m=2$ OLR) but would be more likely to form locked in a more
distant resonance such as the 2:1
mean-motion resonance (at $r=1.6$, as are a number of the
extrasolar planets) or in between, as is Saturn with Jupiter.

\acknowledgments
We thank F.~Masset for his help and for providing the code. We
also thank an unamed referee for considerable help in making the
theory section of the paper stronger by pointing out a number key
conceptual issues. Support for this work was provided by NSF
grants AST-9702484, AST-0098442, NASA grant NAG5-8428, DOE grant
DE-FG02-00ER54600, and the Laboratory for Laser Energetics.
This material is based upon work supported by the National Aeronautics
and Space Admininstration under Grant No. NAGxxxx
issued through the Origins of Solar Systems Program.
This research was supported in part
by the National Science Foundation to the KITP
under Grant No. PHY99-07949.

\clearpage

\smallskip

\clearpage


%
%
%


\clearpage

\begin{deluxetable}{lcrrllll}
\tablecaption{List of numerical simulations \label{tab:run}}
\tablewidth{0pt}
\tablehead{
\colhead{\#} &
\colhead{$q/10^{-3}$} &
\colhead{${\cal R}$}&
\colhead{$t_{end} $} &
\colhead{$\Sigma_{200}$ } &
\colhead{$\Delta_{200}$} &
\colhead{$\Sigma_{1000}$} &
\colhead{$\Delta_{1000}$ }
}
\startdata
1  & 2   & $          10^{4}$& 1000  & $5.0\times 10^{-7}$ & 0.20  & $3.5\times 10^{-7}$  &  0.25  \\
2  & 2   & $2  \times 10^{4}$& 1000  & $1.7\times 10^{-7}$ & 0.22  & $1.1\times 10^{-7}$  &  0.27  \\
3  & 2   & $2.5\times 10^{4}$& 1000  & $1.1\times 10^{-7}$ & 0.22  & $7.0\times 10^{-8}$  &  0.28  \\
4  & 2   & $5  \times 10^{4}$& 1000  & $3.5\times 10^{-8}$ & 0.23  & $1.9\times 10^{-8}$  &  0.30  \\
5  & 2   & $          10^{5}$& 1000  & $1.1\times 10^{-8}$ & 0.27  & $4.2\times 10^{-9}$  &  0.35  \\
6  & 2   & $2  \times 10^{5}$& 2000  & $1.5\times 10^{-8}$ & 0.27  & $1.1\times 10^{-9}$  &  0.35  \\
7  & 2   & $          10^{6}$& 1000  & $4.0\times 10^{-9}$ & 0.31  & $3.5\times 10^{-10}$ &  0.40  \\
8  & 2   & $2  \times 10^{6}$& 2000  & $1.5\times 10^{-8}$ & 0.28  & $6.0\times 10^{-11}$ &  0.42  \\
9  & 2   & $          10^{7}$& 1000  & $3.9\times 10^{-9}$ & 0.32  & $1.5\times 10^{-10}$ &  0.44  \\
10 & 1   & $2  \times 10^{6}$& 1000  & $7.0\times 10^{-9}$ & 0.25  & $2.1\times 10^{-10}$ &  0.33  \\
11 & 0.5 & $2  \times 10^{6}$& 1000  & $8.2\times 10^{-8}$ & 0.17  & $1.5\times 10^{-9 }$ &  0.21  \\
12 & 0.1 & $2  \times 10^{6}$& 1000  & $4.0\times 10^{-7}$ & 0.11  & $5.0\times 10^{-9}$  &  0.16  \\
\enddata
\tablecomments{
Runs listed for different planet mass ratios $q$ and
Reynolds number, ${\cal R}$.
The simulations were run for $t_{end}$ planetary orbital periods.
The densities $\Sigma_{200}, \Sigma_{1000}$ are the azimuthally averaged gas
surface densities at the location of the planet at times $t=200$ and 1000,
respectively.
$\Delta_{200},\Delta_{1000}$ are the distances between the planet
and the edge of the disk (as defined in section 4) at times $t=200$ and 1000,
respectively.
}
\end{deluxetable}

\end{document}